\documentstyle[epsfig,aps]{revtex}
\begin{document}

\title
{Evaporation   residue  cross-sections  as  a  probe  for  nuclear
dissipation in the fission channel of a hot rotating nucleus}

\author{     Gargi    Chaudhuri    \thanks{Electronic    address:
gargi@veccal.ernet.in} and Santanu Pal\thanks{Electronic address:
santanu@veccal.ernet.in}}

\address{Variable  Energy  Cyclotron  Centre,  1/AF Bidhan Nagar,
Kolkata 700 064, India}
\date{\today }
\maketitle
\begin{abstract}

Evaporation  residue cross-sections are calculated in a dynamical
description of nuclear fission in the framework of  the  Langevin
equation coupled with statistical evaporation of light particles.
A  theoretical  model  of  one-body  nuclear  friction  which was
developed earlier, namely the  chaos-weighted  wall  formula,  is
used   in  this  calculation  for  the  $^{224}$Th  nucleus.  The
evaporation residue cross-section is found to be  very  sensitive
to  the  choice of nuclear friction. The present results indicate
that the chaotic nature of the  single-particle  dynamics  within
the  nuclear  volume  may  provide  an explanation for the strong
shape-dependence of nuclear friction which is usually required to
fit experimental data.

\end{abstract}

\pacs{  PACS numbers: 05.40.Jc Brownian motion, 24.60.Lz Chaos in
nuclear  systems,  25.70.Gh Compound nucleus, 25.70.Jj Fusion and
fusion-fission reactions}

\eject

\noindent {\Large {\bf 1 Introduction}}\\

Fission  of  highly excited compound nuclei produced in heavy ion
induced  fusion reactions has evoked considerable interest in the
recent years mainly due to the fact that it cannot  be  accounted
for  by  the  statistical  model  of Bohr and Wheeler for nuclear
fission \cite{bw}. In particular, it is now established that  the
multiplicities  of  neutrons, light charged particles and photons
emitted by a hot compound nucleus  are  much  higher  than  those
predicted by the statistical model \cite{thoe}. This implies that
the  fission  life time is considerably underestimated when using
the Bohr-Wheeler description of  nuclear  fission.  Consequently,
dissipative  dynamical  models for fission of excited nuclei were
developed following the original work of Kramers  who  considered
fission dynamics to be analogous to that of diffusion of Brownian
particles over a barrier \cite{kram}. These dynamical models were
found   to   be   successful  in  reproducing  a  large  body  of
experimentally measured multiplicity data\cite{ab1,fro1}.\\

Dynamical  effects  are  also  found  essential  to calculate the
evaporation residue  cross-section  of  highly  excited  compound
nuclei  \cite{fabris}.  Specifically,  the  decay  width  for  an
overdamped system was subsequently found to be  more  appropriate
for  nuclear  fission and this fission width is generally used in
the description of the statistical decay of an excited  compound
nucleus  \cite{back,dioszegi}.  It turns out that the evaporation
residue cross-section depends strongly on  the  strength  of  the
nuclear  dissipation  whenever it is a very small fraction of the
total fusion cross-section.  In  fact,  the  evaporation  residue
cross-section  in such cases can serve the purpose of a sensitive
probe for nuclear dissipation.\\

The  dissipative  force  in the dynamics of fission arises due to
the interaction between the large  number  of  intrinsic  nuclear
degrees  of  freedom and the few collective or fission degrees of
freedom. The strength of the dissipative force is usually treated
as an adjustable parameter in order to fit the experimental data.
Fr\"obrich  et  al.  \cite{fro2}  obtained   a   phenomenological
shape-dependent  nuclear  dissipation which was able to reproduce
the fission  probability  and  prescission  neutron  multiplicity
excitation  functions  for  a  number  of  compound  nuclei.  The
magnitude of this phenomenological dissipation was  found  to  be
very  small  for  compact  shapes  of  the  nucleus, but a strong
increase in its value was needed for  large  deformations.  In  a
recent  work, Di\'oszegi et al. \cite{dioszegi} have analyzed the
$\gamma$  as  well  as  neutron  multiplicities  and  evaporation
residue  cross-section  of $^{224}$Th and have concluded that the
experimental data can  be  fitted  equally  well  with  either  a
temperature   or  a  deformation-dependent  nuclear  dissipation.
Interestingly,   the   deformation-dependence   of   the    above
dissipation  also corresponds to a lower value of the strength of
the dissipation inside the saddle and a higher value outside  the
saddle,  similar  to  the  phenomenological  dissipation  of ref.
\cite{fro2} mentioned in the above.\\

A  theoretical  model  for  nuclear  dissipation, namely the wall
formula, was developed long ago by Blocki et al. \cite{blocki} in
a simple classical picture of one-body dissipation. However,  the
strength  of  the  wall  dissipation was found to be considerably
higher than that required to fit experimental data.  It  is  only
recently  that a modification has been incorporated into the wall
formula which resulted in a shape-dependent reduction  factor  in
the   strength   of  the  friction\cite{pal1}.  The  modification
essentially arose out of a  closer  examination  of  one  crucial
assumption  of  the  wall formula concerning the randomization of
particle motion within the nuclear volume. It was assumed in  the
original   wall   formula  that  the  particle  motion  is  fully
randomized. This full randomization assumption is relaxed in  the
modified  wall  formula in order to make it applicable to systems
with partly randomized or chaotic single-particle motion. In what
follows, we shall use the  term  ``chaos-weighted  wall  formula"
(CWWF)  for  this modified dissipation in order to distinguish it
from the original wall formula (WF) dissipation. As was shown  in
ref.\cite{pal1}, the CWWF dissipation coefficient ${\eta_{cwwf}}$
will be given as

\begin{eqnarray}
\label{fric}
\eta_{cwwf} = \mu \eta_{wf},
\label{(1)}
\end{eqnarray}

\noindent  where  ${\eta_{wf}}$  is  the  dissipation or friction
coefficient  as  was  given  by   the   original   wall   formula
\cite{blocki}  and  ${\mu}$ is a measure of chaos (chaoticity) in
the single-particle motion and depends on the instantaneous shape
of the nucleus. The value of chaoticity $\mu$ changes from 0 to 1
as the nucleus  evolves  from  a  spherical  shape  to  a  highly
deformed  one. The CWWF dissipation is thus much smaller than the
WF dissipation for  compact  nuclear  shapes  while  they  become
closer  at large deformations. It is worthwhile to note here that
the above  shape-dependent  CWWF  has  features  similar  to  the
empirical dissipations discussed in  the  last  paragraph.  In  a
recent   work,   we  have  shown  that  the  prescission  neutron
multiplicity and fission  probability  calculated  from  Langevin
dynamics  using  the chaos-weighted wall dissipation agree fairly
well with the experimental data for a number  of  heavy  compound
nuclei  ($A  \sim  200$) over a wide range of excitation energies
\cite{gargi1}. This strongly suggests  that  the  shape-dependent
CWWF can be  considered  as  a  suitable  theoretical  model  for
one-body  dissipation in nuclear fission. Further, a lack of full
randomization or chaos in the single-particle motion can  provide
a   physical   explanation  for  the  empirical  requirement  for
reduction in strength of friction for compact nuclear  shapes  in
order to fit experimental data.\\

In  the  present  work,  we  shall employ the chaos-weighted wall
formula  dissipation  to  calculate   the   evaporation   residue
excitation   function   for  the  $^{224}$Th  nucleus.  Our  main
motivation here will be to put the chaos-weighted wall formula to
a further test and verify to what extent it can account  for  the
experimental  evaporation  residue data which is a very sensitive
probe for nuclear dissipation. In our calculation, we shall first
assume  that  a  compound  nucleus  is  formed  when a projectile
nucleus completely fuses with a target nucleus  in  a  heavy  ion
collision. Processes such as fast fission  or  quasifission  thus
cannot  be  described  in  the  model  considered  here. We shall
describe the fission dynamics of  the  compound  nucleus  by  the
Langevin  equation  while  the  light  particles and photons will
undergo statistical  emission.  The  Langevin  equation  will  be
solved  by  coupling it with particle and $\gamma$ evaporation at
each  step  of  its  time  evolution.  The  prescission  particle
multiplicity and fission probability will be obtained by sampling
over a large number of Langevin trajectories. The chaos- weighted
wall  friction  coefficient  is  obtained  following  a  specific
procedure  \cite{blocki2}  which  explicitly  considers  particle
dynamics  in  phase  space  in  order to calculate the chaoticity
factor $\mu$ of eq.(1).  There  is  no  free  parameter  in  this
calculation  of  friction.  The  other  input  parameters for the
dynamical calculation are obtained from standard nuclear  models.
Calculation  will be performed at a number of excitation energies
for $^{224}$Th formed in the $^{16}$O+$^{208}$Pb system. We  have
chosen  this  system  essentially  because of the availability of
experimental data on both  evaporation  residue  and  prescission
neutron  multiplicity  covering  the  same  range  of  excitation
energies and the fact that earlier analyses  of  the  evaporation
residue excitation function have already indicated the need for a
dynamical     model     for     fission     of    this    nucleus
\cite{rossner,morton,brinkmann}.\\

In the following section, we shall briefly describe the dynamical
model along with the necessary input as used in the present work.
The  details  of  the  calculation  will  also be given here. The
calculated   evaporation    residue   excitation   function   and
prescission neutron multiplicities  will  be  compared  with  the
experimental  values in sect.3. A summary of the results along
with the conclusions can be found in the last section.\\

\noindent  {\Large {\bf 2 Langevin description of fission}} \\

An  appropriate  set  of  collective  coordinates to describe the
fission degree of freedom consists of the shape parameters  $c,h$
and  $\alpha$ as was suggested by Brack {\it et al.} \cite{brack}
and we shall employ them in the present calculation.  We  further
simplify  our  calculation  by considering only symmetric fission
($\alpha=0$)  since  the  compound  nucleus  $^{224}$Th  is  much
heavier than the Businero-Gallone transition point. The potential
landscape   in   $(c,h)$   coordinates   is  generated  from  the
finite-range liquid drop model \cite{sierk1} where  we  calculate
the  generalized  nuclear  energy  by  double folding the uniform
density with a  Yukawa-plus-exponential  potential.  The  Coulomb
energy is obtained by double folding another Yukawa function with
the  density distribution. We shall further assume in the present
work that fission would proceed along the valley of the potential
energy  landscape.  Consequently  we  shall  use   an   effective
one-dimensional  potential in the Langevin equation which will be
defined as $V(c)=V(c,h)$ {\it at valley}. This  will  reduce  the
problem  to  one  dimension in order to simplify the computation.
The Langevin equations  in  one  dimension  will  thus  be  given
\cite{wada} as

\begin{eqnarray}
\frac{dp}{dt}   &=&
-\frac{p^2}{2}   \frac{\partial}{\partial   c}\left({1  \over  m}
\right) -
   \frac{\partial F}{\partial c} - \eta \dot c + R(t), \nonumber\\
\frac{dc}{dt} &=& \frac{p}{m} .
\label{(2)}
\end{eqnarray}

\noindent   The   shape-dependent   collective  inertia  and  the
dissipation coefficients in the above equations  are  denoted  by
$m$ and $\eta$ respectively. $F$ is the free energy of the system
while  $R(t)$  represents  the  random  part  of  the interaction
between the fission degree of freedom and the rest of the nuclear
degrees of freedom considered collectively as a thermal  bath  in
the  present  picture.  The  collective  inertia,  $m$,  will  be
obtained by making the Werner-Wheeler approximation \cite{davies}
assuming an incompressible irrotational flow. The  driving  force
in  a thermodynamic system should be derived from its free energy
which  we  will  calculate   considering   the   nucleus   as   a
noninteracting  Fermi  gas  \cite{fro2}. The instantaneous random
force $R(t)$ is assumed  to  have  a  stochastic  nature  with  a
Gaussian  distribution  whose  average  is  zero  \cite{ab1}. The
strength  of  the  random  force  will  be  determined   by   the
dissipation   coefficient   through  the  fluctuation-dissipation
theorem. The various  input  quantities  are  described  in  some
detail in a recent publication \cite{gargi1}.\\ \\

\noindent{\large {\bf 2.1 Nuclear dissipation}}\\

We   shall   use   the  chaos-weighted  one-body  wall-and-window
dissipation in the present calculation. We shall beiefly describe
here the essential features of this dissipation, the  details  of
which    may   be   found   elsewhere   \cite{gargi1}.   In   the
wall-and-window  model  of  one-body  dissipation,   the   window
friction  is  expected  to be effective after a neck is formed in
the nuclear system \cite{sierk2}. Further, the radius of the neck
connecting the two future fragments should be sufficiently narrow
in order to make the energy transfer irreversible.  It  therefore
appears that the window friction should be very nominal when neck
formation  just  begins. Its strength should increase as the neck
becomes narrower reaching  its  classical  value  when  the  neck
radius  becomes  much  smaller  than  the  typical  radii  of the
fragments. We shall approximately describe the above scenario  by
defining   a   transition   point  $c_{win}$  in  the  elongation
coordinate at which the window friction will  be  turned  on.  We
shall also assume that the compound nucleus evolves into a binary
system  beyond $c_{win}$ and accordingly correction terms for the
motions of the centers of mass of the two halves will be  applied
to  the  wall  formula for $c>c_{win}$ \cite{sierk2}. However, it
may be noted that the window dissipation and the center  of  mass
motion  correction  tend  to  cancel  each  other to some extent.
Consequently, the resulting wall-and-window friction is not  very
sensitive  to the choice of the transition point. We shall choose
a value for $c_{win}$ at which the nucleus has a binary shape and
the neck radius is half of the radius of either of  the  would-be
fragments.\\

The  wall-and-window dissipation and its chaos-weighted version
will thus be given as

\begin{eqnarray}
\eta_{wf}(c < c_{win})= \eta_{wall}(c < c_{win}),
\label{(3)}
\end{eqnarray}

\noindent and

\begin{eqnarray}
\eta_{wf}(c \ge c_{win})= \eta_{wall}(c \ge c_{win}) +
\eta_{win}(c \ge c_{win}),
\label{(4)}
\end{eqnarray}

\noindent while

\begin{eqnarray}
\eta_{cwwf}(c < c_{win})= \mu (c) \eta_{wall}(c < c_{win}),
\label{(5)}
\end{eqnarray}

\noindent and

\begin{eqnarray}
\eta_{cwwf}(c \ge c_{win})= \mu(c) \eta_{wall}(c \ge c_{win}) +
\eta_{win}(c \ge c_{win}).
\label{(6)}
\end{eqnarray}

The   detailed  expressions for the wall-and-window frictions can
be found in ref. \cite{sierk2}.\\

\begin{figure}[htb]
\centering
\epsfig{figure=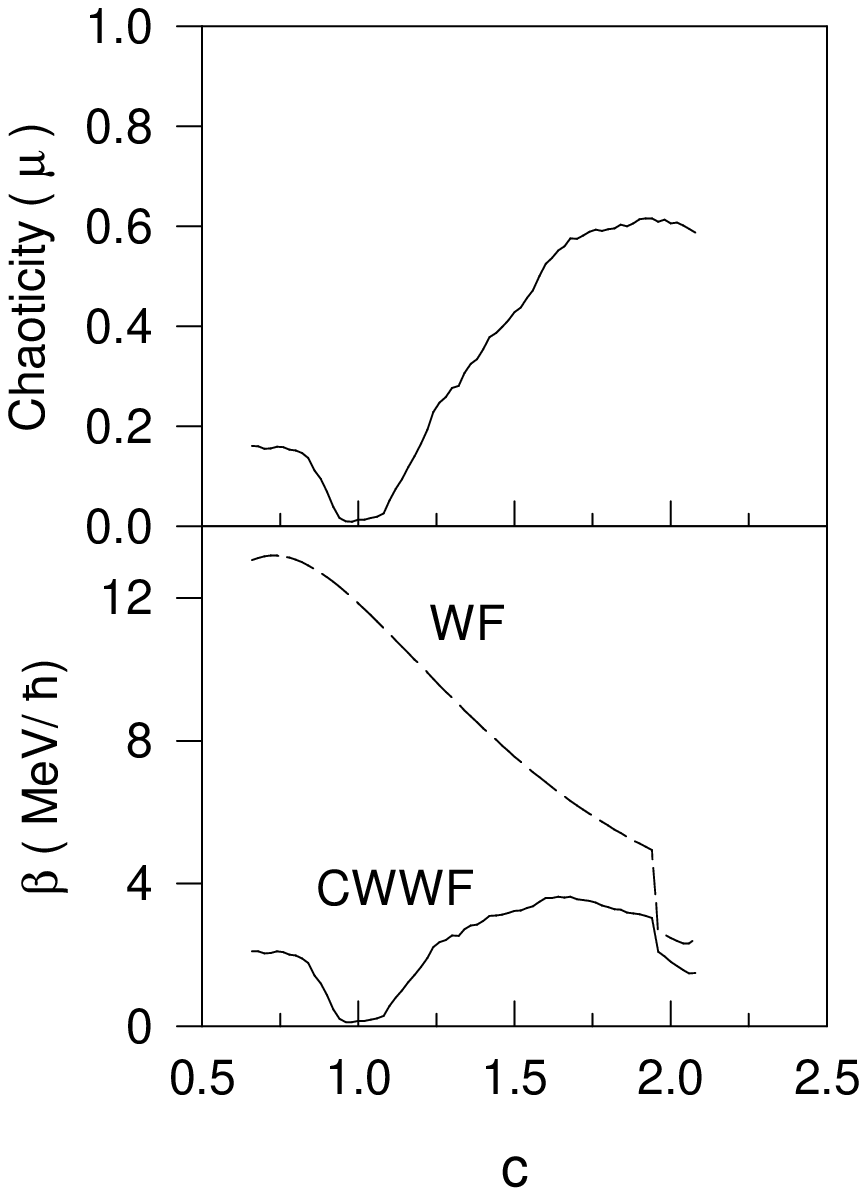}
\caption{\label{fig1}Shape  dependence  of  chaoticity $\mu$, the
reduction  factor  in   chaos-weighted   wall   friction   (upper
panel),and   the  reduced  friction  $\beta$  (lower  panel)  for
$^{224}$Th.}
\end{figure}

The  chaoticity  ${\mu(c)}$,  introduced  in the earlier section,
depends on the instantaneous shape of the nucleus \cite{pal1}. In
a classical picture, this will be given as the  average  fraction
of  the nucleon trajectories within the nucleus which are chaotic
when the sampling is done uniformly over the nuclear surface. The
value of the chaoticity for a given nuclear shape is evaluated by
sampling over a large number of  classical  nucleon  trajectories
while  each  trajectory is identified either as a regular or as a
chaotic one by considering the magnitude of its Lyapunov exponent
and the nature of its variation  with  time  \cite{blocki2}.  The
shape-dependence  of  the  chaoticity, thus obtained, is shown in
fig.\ref{fig1}. Further, defining a  quantity  $\beta  (c)=  \eta
(c)/ m(c)$ as the reduced dissipation coefficient, its dependence
on  the elongation coordinate is also shown in fig.\ref{fig1} for
the $^{224}$Th nucleus. It can be immediately  noticed  that  the
CWWF  is  strongly  suppressed compared to the WF dissipation for
near-spherical shapes ($c \sim 1$) and this can be  qualitatively
understood  as  follows.  A  particle  moving in a spherical mean
field represents a typical integrable system and its dynamics  is
completely  regular.  When  the boundary of the mean field is set
into motion (as in fission), the energy gained by the particle at
one instant as a result of a collision with the  moving  boundary
is  eventually  fed  back to the boundary motion in the course of
later collisions. An integrable system  thus  becomes  completely
nondissipative   in   this   picture  resulting  in  a  vanishing
dissipation coefficient. It may be noted that the suppression  of
friction  strength  in  CWWF  is  qualitatively  similar  to  the
shape-dependent frictions found empirically  \cite{dioszegi,fro2}
to  fit experimental data. The considerations of chaos (or rather
lack of it) in  particle  motion  can  thus  provide  a  physical
explanation  for  the reduction in friction strength required for
compact shapes of the compound nucleus.\\

The strong shape-dependence of the CWWF can have some interesting
consequences.  In  a dynamical description of fission, a compound
nucleus spends most of its time in undergoing shape  oscillations
in  the  vicinity  of its ground state shape before it eventually
crosses the saddle and proceeds towards the scission point. Since
the  spin  of  a compound nucleus formed at a small excitation is
also small, its ground state shape is  nearly  spherical  and  in
this region the CWWF friction is also small.  Conversely,  higher
spin values are mostly populated in  a  highly  excited  compound
nucleus making its ground state shape highly deformed and thus it
experiences  a  strong  CWWF  friction.  Therefore, if one uses a
shape-independent friction in a dynamical model of  fission,  its
strength  has to increase with increasing temperature in order to
give  an  equivalent  description  to  that   provided   by   the
temperature-independent  but  shape-dependent  CWWF  friction. In
fact,  it  was  observed   in   ref.   \cite{dioszegi}   that   a
shape-dependent  friction fits the experimental data equally well
to that achieved  by  a  strong  temperature-dependent  friction.
Since  there  is a physical justification for shape-dependence in
nuclear friction from chaos considerations, it  is  quite  likely
that   the   above  strong  temperature-dependence,  at  least  a
substantial part of it, is of dynamical origin  as  explained  in
the  above  and  thus  is  an  artifact  arising  out  of using a
shape-independent friction.\\

It  must  be  pointed  out,  however,  that  one   would expect a
temperature-dependence   of   nuclear   friction   from   general
considerations such as larger phase space becoming accessable for
particle-hole   excitations   at   higher   temperatures.   In  a
microscopic model of  nuclear  friction  using  nuclear  response
function,  Hofmann  {\it  et  al.} \cite{hofmann} have obtained a
nuclear friction which depends  upon  temperature  as  $0.6T^{2}$
(leading   term).   This  may  be  compared  with  the  empirical
temperature-dependent  term  of  $3T^{2}$  which  was  found   in
ref.\cite{dioszegi}.  It  therefore  appears  that  only  a small
fraction of the empirical temperature-dependence can be accounted
for by the inherent temperature-dependence  of  nuclear  friction
while  the rest of it has a dynamical origin as we have discussed
in the above.\\

In  the  present  work,  we  shall  not  consider  any  empirical
temperature-dependence of the CWWF or WF frictions  in  order  to
study solely the effects of shape-dependence. In what follows, we
shall  use both the WF and CWWF dissipations in a dynamical model
of fission and shall investigate the effect of the  reduction  in
the CWWF strength on the evaporation residue cross-section.\\

\noindent{\large {\bf 2.2 Dynamical model calculation}}\\

In   the  dynamical  model  calculation,  the  initial  spin  and
excitation energy of the compound nucleus is determined from  the
entrance  channel  specifications  in  the  following manner. The
fusion cross-section of the target and projectile in the entrance
channel usually obeys the following spin distribution

\begin{eqnarray}
\frac{d \sigma(l)}{dl} = \frac{\pi}{k^2}
\frac{(2l+1)}{1+\exp \frac{(l-l_{c})}{\delta l}}
\label{(7)}
\end{eqnarray}

\noindent  where  we  shall  obtain  the  parameters  $l_{c}$ and
$\delta l$ by fitting the experimental fusion cross-sections. The
initial spin of the compound nucleus in our calculation  will  be
obtained  by  sampling  the above spin distribution function. The
total excitation energy ($E^{*}$) of the compound nucleus can  be
obtained  from  the  beam  energy  of  the projectile, and energy
conservation in the form

\begin{equation}
E^{*}=E_{int}+V(c)+p^{2}/2m    \label{(8)}
\end{equation}

\noindent gives the intrinsic excitation energy $E_{int}$ and the
corresponding nuclear temperature $T=(E_{int}/a)^{1/2}$ where $a$
is the nuclear level density parameter. The centrifugal potential
is included in $V(c)$ in the above equation.

We  shall  use  the  following  level  density  parameter  due to
Ignatyuk {\it et. al.} \cite{igna} which incorporates the nuclear
shell structure at low excitation energy and goes smoothly to the
liquid drop behavior at high excitation energy,

\begin{eqnarray}
a(E_{int})=\bar{a}(1+\frac{f(E_{int})}{E_{int}} {\delta M} ),
\label{(9)}
\end{eqnarray}

\noindent with

\begin{eqnarray}
f(E_{int})=1-exp(-E_{int}/E_{D})
\nonumber
\end{eqnarray}

\noindent where  $\bar{a}$  is  the  liquid  drop  level  density
parameter, $E_{D}$ determines the rate at which the shell effects
disappear at high  excitations,  and  $\delta  M$  is  the  shell
correction  given  by the difference between the experimental and
liquid drop  masses,  $(\delta  M=M_{exp}-M_{LDM}  )$.  We  shall
further   use  the  shape-dependent  liquid  drop  level  density
parameter given as \cite{balian},

\begin{eqnarray}
\bar{a}(c)=a_{v}A+a_{s}A^{\frac{2}{3}}B_{s}(c)
\label{(10)}
\end{eqnarray}

\noindent  where we choose the values for the parameters $a_{v}$,
$a_{s}$ and the  dimensionless  surface  area  $B_{s}$  following
ref.\cite{fro2}.\\

In  order  to solve the Langevin equations, the initial values of
the coordinates and momenta $(c,p)$  of  the  fission  degree  of
freedom  are  obtained from sampling random numbers following the
Maxwell-Boltzmann distribution. The Langevin  equations  (eq.(2))
are  subsequently  numerically integrated following the procedure
outlined in ref.\cite{ab1}. At each time step of  integration  of
the  Langevin equations, particle (neutron, proton and alpha) and
giant dipole $\gamma$ evaporation will be considered following  a
Monte Carlo sampling technique \cite{fro2}. For this purpose, the
particle  and  $\gamma$  decay  widths  are  calculated using the
inverse cross-section formula as given in ref.\cite{fro1}.  After
each  particle  emission,  the  potential energy landscape of the
parent nucleus is replaced by that of the daughter  nucleus.  The
intrinsic  excitation  energy,  mass  and  spin  of  the compound
nucleus are also recalculated after each emission.  The  spin  of
the  compound  nucleus  is  reduced only in an approximate way by
assuming that each neutron, proton or  a  $\gamma$  carries  away
$1\hbar$ angular momentum while that carried by an alpha particle
is $2\hbar$. However,  only  neutron  emission  is  found  to  be
relevant for the yield of the residue cross-section.\\

A  Langevin trajectory will be considered as undergone fission if
it reaches the scission point ($c_{sci}$) in course of  its  time
evolution.  Alternately  it  will  be  counted  as an evaporation
residue event if the intrinsic excitation energy becomes  smaller
than  either  the  fission  barrier  or  the  binding energy of a
particle. The calculation proceeds  until  the  compound  nucleus
undergoes   fission  or  becomes  an  evaporation  residue.  This
calculation  is  repeated  for  a  large   number   of   Langevin
trajectories and the evaporation residue formation probability is
obtained  as the fraction of the trajectories which have ended up
as evaporation residues. The evaporation residue cross-section is
subsequently obtained by multiplying the experimental  value  for
fusion  cross-section  in  the  entrance  channel  with the above
formation probability of the evaporation residue. Similarly,  the
average number of particles (neutrons, protons or alphas) emitted
in the fission events will give the required prescission particle
multiplicities.\\

Following  the  fission  dynamics  through  the Langevin equation
during the entire life time of a  compound  nucleus  can  however
take   an   extremely  long  computer  time.  As  an  alternative
procedure, we shall first follow the time evolution of a compound
nucleus according to the Langevin equations  for  a  sufficiently
long period during which a steady flow across the fission barrier
is  established.  Beyond  this  period,  a  statistical model for
compound nucleus decay is expected to be a equally valid and more
economical in terms of computation.  We  shall  therefore  switch
over to a statistical model description after the fission process
reaches  the  stationary  regime.  This  combined  dynamical  and
statistical model,  first  proposed  by  Mavlitov  {\it  et  al.}
\cite{mavli}  , however, requires the fission width along with the
particle and $\gamma$ widths in the  statistical  branch  of  the
calculation. This fission width should be the stationary limit of
the fission rate as determined by the Langevin equation. However,
it  is  not possible to obtain this fission rate analytically for
the strongly shape-dependent CWWF and WF  disipations.  We  shall
therefore  use  a  suitable  parametric  form  of the numerically
obtained stationary fission widths using the CWWF (and  also  WF)
dissipations  in  order  to use them in the statistical branch of
our calculation. The  details  of  this  procedure  is  given  in
ref.\cite{gargi2}  which  we  shall  follow  to calculate all the
required fission widths for the present work.\\

\noindent{\Large {\bf 3 Results}}\\

\begin{figure}[htb]
\centering
\epsfig{figure=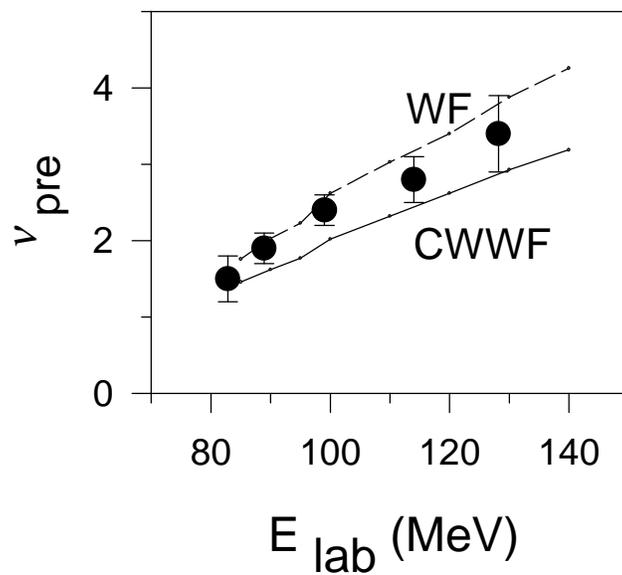}
\caption{\label{fig2}Prescission       neutron       multiplicity
($\nu_{pre}$) excitation  function  calculated  with  WF  (dashed
line)   and   CWWF   (full   line)  frictions  for  the  reaction
$^{16}$O+$^{208}$Pb. The  experimental  points  (dots)  are  also
shown.}
\end{figure}

We  have  calculated the prescission neutron multiplicity and the
evaporation residue (ER) cross-section for the  compound  nucleus
$^{224}$Th  when  it  is  formed  in  the  fusion  of an incident
$^{16}$O  nucleus  with  a   $^{208}$Pb   target   nucleus.   The
calculation is done at a number of incident energies in the range
of 80 MeV to 140 MeV using both the WF and the CWWF dissipations.
Figure   \ref{fig2}  shows  the  calculated  prescission  neutron
multiplicity along with  the  experimental  data  \cite{rossner}.
Both the WF and CWWF predictions for multiplicity are quite close
to   the   experimental   values  and  this  shows  that  neutron
multiplicity is not very sensitive to the dissipation in  fission
in  the energy range under consideration. It must be pointed out,
however, that the CWWF predictions for neutron  multiplicity  are
closer  to  experimental  data  compared to those from WF at much
higher excitations of the compound nucleus\cite{gargi1}.\\

It  may  be  mentioned at this point that though we calculate the
number  of  prescission protons, alphas and GDR $\gamma$'s, we do
not compare them with experimental data because these numbers are
rather  small with large statistical uncertainties in the present
work.  In  order  to  obtain  the energy spectrum of the $\gamma$
multiplicity  with  a   reasonable   statistical   accuracy,   in
particular,  it  is necessary to perform computation using a much
larger ensemble of trajectories than the one (comprising of 20000
trajectories) used here. This puts a severe  demand  on  computer
time making such computations impractical at present. However, an
alternative  approach  would be to make use of the time-dependent
fission widths in a full statistical model calulation of compound
nucleus decay. This calculation would be  much  faster  than  the
present   Langevin   dynamical   model   calculation  though  the
time-dependent fission widths would be required as input to  this
statistical   model   calculation.   We   plan  to  perform  such
calculations in future.\\

\begin{figure}[htb]
\centering
\epsfig{figure=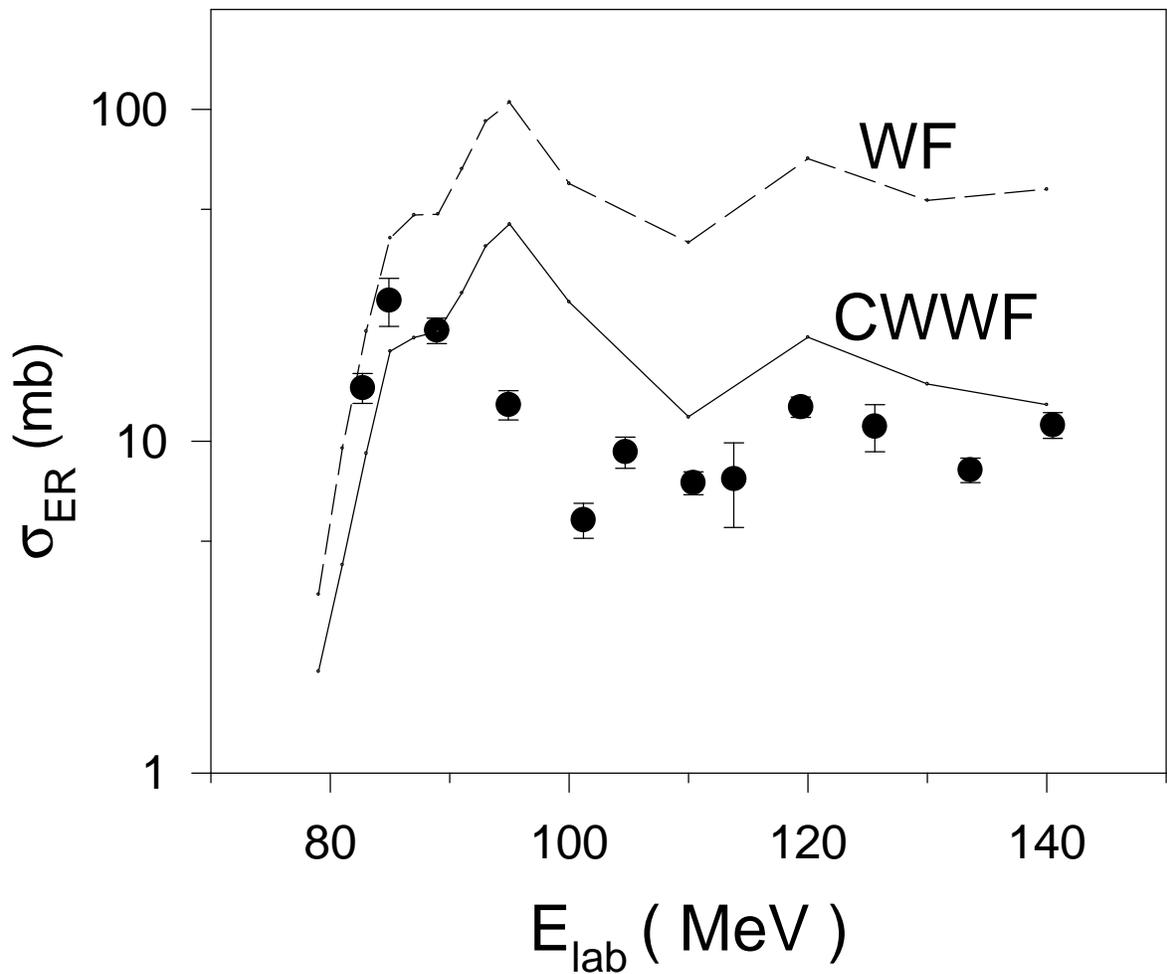}
\caption{\label{fig3}Evaporation residue cross-section excitation
function  calculated  with  WF (dashed line) and CWWF (full line)
frictions for the reaction as in fig.2. The  experimental  points
(dots) are also shown.}
\end{figure}

We  shall  next  consider  the results of the evaporation residue
calculation. Figuer  \ref{fig3}  shows  the  evaporation  residue
excitation  functions  calculated  using  both  the  WF  and CWWF
dissipations. The  experimental  values  of  evaporation  residue
cross-section  are also shown in this figure \cite{brinkmann}. We
first note that the calculated evaporation residue  cross-section
is  very  sensitive  to  the dissipation in the fission degree of
freedom.  The  WF  predictions  are  a  few times (typically 2-5)
larger than those obtained with the  CWWF  dissipation.  Next  we
make the important observation that the CWWF predicted excitation
function  is  much  closer  to  the experimental values than that
obtained with the WF dissipation. This observation clearly  shows
that  the  chaos-weighted  factor in CWWF changes its strength in
the right direction. We must take note of the fact, however, that
the CWWF still considerably overestimates  the  ER cross-section.
Since  the  present  dynamical  calculation  considers  only  one
(elongation)  fission  degree  of  freedom,  it  is expected that
inclusion of the neck degree of freedom will increase the fission
probability  \cite{wada2}  further  and  hence  reduce   the   ER
cross-section.  We  plan  to extend our work in this direction in
future. We further observe that  while  a  peak  appears  in  the
experimental  excitation  function  at  about 85 MeV, the same is
shifted by 10 MeV in the calculated results. We do not  have  any
explanation  for  this discrepancy except pointing out that there
is no free parameter in our calculation  and  thus  no  parameter
tuning  has  been  attempted in order to fit experimental data. A
similar  shift  has  also  been  observed  in  an  earlier   work
\cite{dioszegi}.\\

The  structure of the evaporation residue excitation function can
also reveal certain interesting features.  Since  the  calculated
values  of  the evaporation residue cross-section are obtained as
the product of the fusion cross-section and  the  probability  of
evaporation  residue  formation,  the  initial  rise  of  the  ER
cross-section with beam energy  essentially  reflects  the  steep
rise of fusion cross-section in this energy region \cite{morton}.
At  still  higher  beam  energies,  the  ER cross-section becomes
approximately  stable  which  results  from  a  delicate  balance
between  the increasing trend of the fusion cross-section and the
decreasing trend of the probabilty of ER formation.  Had  the  ER
formation  probability  decreased  at  a  rate  higher than those
obtained  in  the   present   calculation,   the   resulting   ER
cross-section  would  have  decresed  at  higher compound nuclear
excitations.  In  fact,  such  an   observation   was   made   in
ref.\cite{brinkmann}  where  the  ER  cross-section obtained from
standard statistical model calculation was found to decrease very
steeply beyond 100 MeV of beam energy. In order to  explore  this
point  a  little  further,  we  have  calculated  the  excitation
function of the partial width for fission. Since fission can take
place at any stage during neutron (or any other  light  particle)
evaporation, the partial widths are calculated for $^{222}$Th and
$^{223}$Th  as well at excitation energies reduced by the neutron
separation energy  after  each  neutron  emission.  The  compound
nuclear  spin  was  taken  as $l_{c}$ from eq.(7) while only CWWF
dissipation was considered for this calculation.\\

\begin{figure}[htb]
\centering
\epsfig{figure=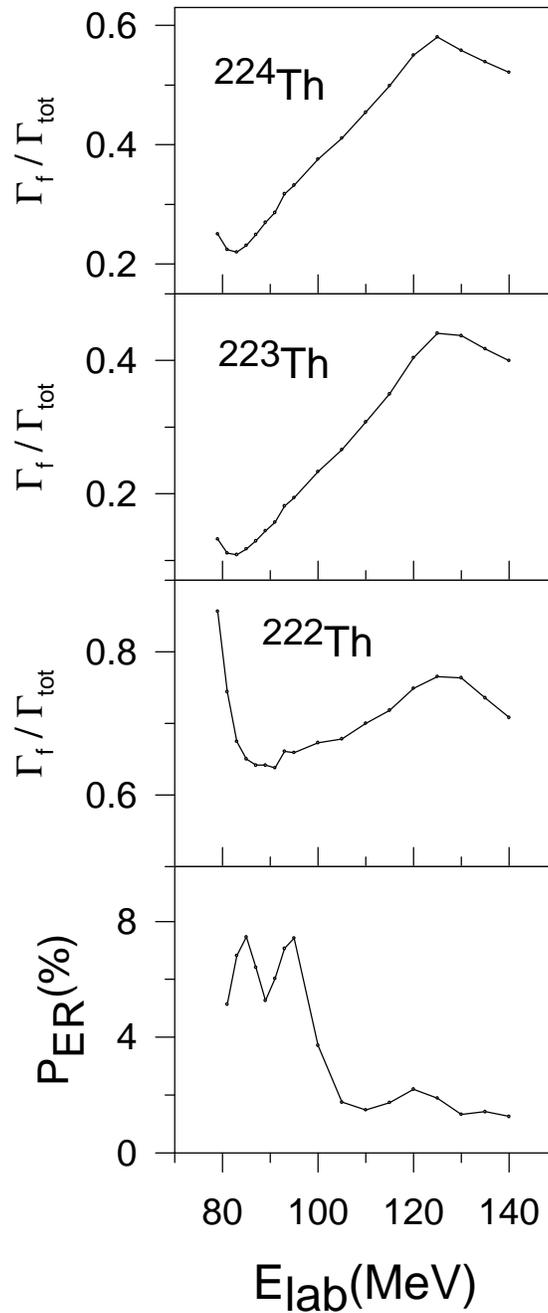}
\caption{\label{fig4}The   top  three  panels  show  the  fission
partial widths for $^{224}$Th,  $^{223}$Th  and  $^{222}$Th  (see
text).  The  total  width  $\Gamma  _{tot}$ includes the neutron,
proton, alpha and $\gamma$ evaporation widths in addition to  the
fission  width. The bottom panel displays the excitation function
of the evaporation residue formation probability for the reaction
as in fig.2}
\end{figure}

The calculated excitation functions of the fission partial widths
are  shown  in  fig.  \ref{fig4}. The calculated values of the ER
formation probability  ($P_{ER}$)  are  also  displayed  in  this
figure. Each partial width excitation function is found to have a
minimum  around  90  MeV  of  beam  energy  after which it starts
increasing till this trend is arrested  and  reversed  at  higher
excitations. Recalling the fact that the above results on partial
widths are only indicative while $P_{ER}$ is obtained from a full
dynamical  calculation,  it is of interest to note that a bump in
the excitation function of $P_{ER}$ also  appears  in  the  above
($\sim$ 90 MeV) energy range. Subsequently, the value of $P_{ER}$
drops  rather  sharply  before  it  settles  to a steady value at
higher excitations. This feature is also complementary to that of
the excitation functions of the partial  widths  of  fission.  We
thus  demonstrate  in a schematic manner how the structure in the
excitation function of the ER cross-section  is  related  to  the
competition between fission and other decay channels at different
stages of fission.\\

\noindent{\Large {\bf 4 Summary and conclusions}}\\

We have applied a theoretical model of one-body nuclear friction,
namely   the   chaos-weighted   wall   formula,  to  a  dynamical
description of compound nuclear decay where fission  is  governed
by the Langevin equation coupled with the statistical evaporation
of  light  particles. We have used both the standard wall formula
and its modified form with the chaos-weighted factor in order  to
calculate  the  prescission  neutron multiplicity and evaporation
residue excitation functions for the $^{224}$Th  nucleus.  Though
the  number of the prescission neutrons calculated with either WF
or CWWF friction are found to be very close to each other in  the
energy range considered, the evaporation residue cross-section is
found  to depend very strongly on the choice of nuclear friction.
The evaporation residue cross-section calculated  with  the  CWWF
friction gives a much better agreement with the experimental data
compared to the WF predictions. This result demonstrates that the
consequences  of  chaos  in particle motion give rise to a strong
suppression of the strength of  the  wall  friction  for  compact
shapes   of   the   compound   nucleus  which  ,in  turn,  brings
theoretically  calculated  evaporation   residue   cross-sections
considerably  closer  to  the experimental values. Thus the chaos
considerations  may  provide  a  plausible  explanation  for  the
shape-dependence  of  the  strength of nuclear friction which was
found \cite{dioszegi,fro2}  to  be  necessary  in  order  to  fit
experimental data.

\eject

\eject

\eject

\eject

\eject

\end{document}